\documentstyle{article}
\begin{document}
\sloppy
\section*{Dynamical zeta functions for Artin's billiard and the
Venkov--Zograf factorization formula}
\begin{center}
Michael Eisele\\
IFF, KFA\\
D-W-5170 J\"ulich, Germany\\
\&\\
Dieter Mayer\\
Institute of Theoretical Physics\\
Technical University\\
D-W-3392 Clausthal-Zellerfeld, Germany\\
\end{center}

\section*{Abstract}
Dynamical zeta functions are expected to relate the Schr\"odinger
operator's spectrum to the periodic orbits of the corresponding
fully chaotic Hamiltonian system. The relationsship is exact
in the case of surfaces of constant negative curvature.
The recently found factorisation of the Selberg zeta function for the modular
surface is known to correspond to a decomposition of the
Schr\"odinger operator's eigenfunctions into two sets obeying different
boundary condition on Artin's billiard. Here we express
zeta functions for Artin's billiard in terms of generalized
transfer operators, providing thereby a new dynamical proof of the above
interpretation of the factorization formula.
This dynamical proof
is then extended to the Artin--Venkov--Zograf formula for
finite coverings of the modular surface.

\section{Introduction}
Dynamical zeta fuctions for flows $\phi_t : M \to M$ were introduced
in the sixties by Smale \cite{r1} after he learned from Sinai about
such an     interpretation
for Selberg's zeta fuction \cite{r2} in terms of the geodesic flow
on a surface of constant negative curvature defined by a discrete group acting
on Poincar\'e's hyperbolic upper half plane \cite{r3},
the so called Fuchsian group. It turned out that
Ruelle's and Bowen's thermodynamic formalism \cite{r4} provides
a new approach for understanding the properties of these functions
completely different from Selberg's original one, which was based   mainly
on his famous trace formula \cite{r5}. In the new approach the
dynamical zeta function gets expressed in tems of Fredholm determinants
of so called transfer operators \cite{r6}. These operators haver their origin
in the transfer matrix method of statistical mechanics and were used
in the early days of the thermodynamic formalism to characterize the ergodic
properties of dynamical systems \cite{r7}.

Since in the case of the geodesic flow on surfaces of constant negative
curvature there exists a close connection between the nontrivial zeros
of the corresponding dynamical zeta function and the spectrum
of the free Schr\"odinger operator on the surface, the transfer
operator, which is a purely classical object, when analytically continued to
complex temperatures, hence determines also the quantum properties of
these systems \cite{r8}, \cite{r9}. One could then expect that something
like this could be true also for general chaotic Hamiltonian systems
at least in the semiclassical limit of the systems' quantization. Indeed
a heuristic approach to this question was recently provided in
\cite{r9} where the quantization condition is again expressed in terms
of a Fredholm determinant of a so called quantum operator, which coincides
with the aforementioned transfer operator at least in the case
of the geodesic flow on a compact surface of constant negative
curvature \cite{r10}.

Not so much is known for such surfaces in case they are not compact,
but have a finite hyperbolic area \cite{r11}. Among them the modular
surface, defined through the action of the modular group $SL(2, {\bf Z})$
on the Poincar\'e half plane ${\bf H} = \{ z = x + iy : y > 0 \}$ plays a
special role: for its geodesic flow the above theory works perfectly
and the dynamical zeta fuction can be expressed again in terms of the
Fredholm determinant of a transfer operator \cite{r11}, providing at the
same time a much easier and shorter approach to Selberg's results for
such noncompact surfaces of constant negative curvature. Furthermore, an
interesting factorization formula of the dynamical zeta function follows
from the transfer operator approach, closely related to the
description of the geodesic flow on
the modular surface in terms of the symbolic
dynamics of Series \cite{r12} resp. Adler and Flatto \cite{r13}. The
conjecture in \cite{r8}, namely, that this factorization corresponds
to a decomposition of the spectrum of the free Schr\"odinger--operator
on the modular surface into even respectively odd eigenfunctions
under reflection of its fundamental domain
on the imaginary axis, was proved recently in \cite{r14} by using
number theoretic methods. A formal argument was given also in \cite{r10}.

In the present paper we give a rigorous proof of
this conjecture based completely on dynamical properties of the flow and its
relation to the Artin billiard. It turns out that our proof shows at the
same time that the factorization of the Selberg
function for the geodesic flow on the modular surface can be interpreted as an
extension of the Artin--Venkov--Zograf factorization \cite{r15} for the
Selberg function for normal subgroups $\Gamma'$ of Fuchsian groups $\Gamma$
in terms of Selberg functions with unitary
representation of this group $\Gamma$. Indeed the modular group
$SL(2, {\bf Z})$ is a normal subgroup of $GL(2,{\bf Z})$
and the Selberg functions of the
Artin billiard with Neumann resp. Dirichlet boundary conditions are just
generalized zeta functions of the latter group corresponding to the two
one dimensional unitary representations of $GL(2, {\bf Z})$
with kernel $SL(2, {\bf Z})$.
We then show how our approach to the factorization formula for the modular
surface can be extended immediately to any finite covering
of the modular surface, defined by a normal subgroup $\Gamma'$ of the modular
group $SL(2, {\bf Z})$, and results in a simple dynamical proof of the
Artin--Venkov--Zograf formula. An interesting case is the principal
congruence subgroup
$\Gamma(2) = \{g \in SL(2, {\bf Z}): g \mbox{ mod } 2 = {\bf 1}\}$
which defines a sixfold covering of the modular surface. Its geodesic
flow is closely related to certain approximate solutions of the so called
mixmaster universe \cite{r16}.

In a first section we determine the symbolic dynamics of the billiard flow
on Artin's billiard from the one determined by Series \cite{r12} resp.
Adler and Flutto \cite{r13} some time ago for the geodesic flow on
the modular surface. This allows us to express the dynamical zeta functions
for the billiard flow in terms of Fredholm determinants of the
generalized transfer operator for the Gau\ss{} map. We show that the product
of these two functions gives exactly the dynamical zeta fuction for the
geodesic flow on the modular surface. Using well--known results of Venkov
this then proves our conjecture relating the factorization to the even
and odd spectrum of the Schr\"odinger operator.
Next we show that our factorization can be regarded also as an
extension of the Artin--Venkov--Zograf formula for subgroups of
Fuchsian groups to the case of the subgroup $SL(2, {\bf Z})$ of the group
$GL(2, {\bf Z})$.

In a last section we argue how our approach can be used to provide a
dynamical proof of the Artin--Venkov--Zograf formula for finite coverings
of the modular surface. These authors used in their work the special
group theoretic structure of the corresponding Fuchsian groups whereas ours
is mainly of dynamical nature.

Work of the authors on this paper started during a common stay
at the Institute for Solid State Physics of Prof. P. Szepfalusy in Budapest
financed through the German-Hungarian exchange program. The results of
this paper were announced by one of us at the workshop ``Symbolic
Dynamics'' at the MSRI at Berkeley (Calif.) in Nov. 1992.

\section{Dynamical zeta function for Artin's billiard}
\subsection{Geodesic flow on the modular surface}
The Poincar\'e half plane ${\bf H} = \{ z = x + iy : y > 0 \}$ is
shown in figure \ref{fullview}. Due to it's metric $(dx^2 + dy^2)/y^2$
the geodesics are circles (or straight lines) orthogonal to the
x--axis. The modular surface is defined by identifying points $z$,
$T z = z+1$ and $Qz = -1/z$. $T$ and $Q$ are the generators of the
group
\begin{equation} \label{psl}
PSL(2, {\bf Z}) = SL(2, {\bf Z}) / \{ +{\bf 1}, -{\bf 1} \}
\end{equation}
acting on ${\bf H}$ as
\begin{equation} \{f:{\bf H} \to {\bf H} | f(z) =
\frac{\alpha z + \beta}{\gamma z + \delta}; \alpha \delta - \beta \gamma = 1;
\alpha, \beta, \gamma, \delta \in {\bf Z}\}\quad .
\end{equation}
The fundamental domain $\{z \in {\bf H}: |z| \ge 1; |\Re z| \le 1/2 \}$
tiles the plane ${\bf H}$ under the action of this group.
Artin's billiard is the part of the fundamental domain with $x > 0$.
At intersections with the boundary of this domain
the geodesic flow of the billiard is reflected while the geodesic flow
of the modular surface is mapped under $Q$ or $T$ or $T^{-1}$ to an
equivalent point of the fundamental domain of the modular group
$SL(2, {\bf Z})$.

The symbolic dynamics of the geodesic flow on the modular surface
was investigated by Adler and Flatto \cite{r13} resp. Series \cite{r12}.
They introduced a
Poincar\'e section, which can be finally induced on $I_2 \times
{\bf Z}_2$.
The coordinates $(\chi ,\psi , \epsilon) \in I_2 \times
{\bf Z}_2$ of this induced Poincar\'e section
are determined from the orbit's $\gamma$
forward and backward intersections  $x_{+\infty}$ and $x_{-\infty}$ with
the $x$-axis.
Immediately before the intersection with the unit circle we have
$x_{+\infty} \in (-1, 1)$ and:
\begin{eqnarray}
\chi &:=& |x_{+\infty}| \in (0, 1)\label{mdef1}\\
\psi &:=& 1 / |x_{-\infty}| \in (0, 1)\label{mdef2}\\
\epsilon &:=& {\rm sgn}(x_{+\infty} - x_{-\infty}) \in \{-1, 1\} \label{mdef3}
\quad .
\end{eqnarray}
In these coordinates the Poincar\'e Map $\tilde{T}$
takes the simple form \cite{r20}:
\begin{equation} \label{ttilde}
\tilde{T}(\chi ,\psi , \epsilon) = (T_G(\chi ), \frac{1}{\psi + [1/\chi]},
-\epsilon)
\end{equation}
where $[x]$ denotes the integer part of $x$ and $T_G$ is the Gau\ss{}
map on the unit interval:
\begin{equation}
T_G(\chi) := 1/\chi \quad {\rm mod} \quad 1 \quad , \chi \not= 0 \quad.
\end{equation}
The $x$-component of the direction of motion changes its sign $\epsilon$ at
every intersection with the unit circle. Thus the number of
these intersections along a periodic orbit is equal to the number of
Poincar\'e mappings.
Enumerating the monotonic branches of the Gau\ss{} map with the natural
numbers one gets a complete and exact symbolic dynamics of the Gau\ss{} map.
This generalizes to a symbolic dynamics of the Poincar\'e map $\tilde{T}$. Thus
all the periodic orbits $\gamma$ of the modular surface are classified.
Their length $l(\gamma)$ was determined by Pollicott \cite{r20}
as a sum over all their intersections with
the Poincar\'e section:
\begin{eqnarray}
l(\gamma) &=& \sum_i r(\chi_i ,\psi_i , \epsilon_i) \label{mdefl}\\
r(\chi ,\psi , \epsilon) &:=& r(\chi) := \ln | T_G^`(\chi)|
\quad .\label{mdefr}
\end{eqnarray}

\subsection{Periodic orbits of the modular surface and the Artin billiard}
Let us denote the periodic orbits of the
geodesic flow on the Artin billiard by $\gamma_b$ and
on the modular surface by $\gamma_m$. We will derive a relationship between
their lengths and numbers.

This relationship depends on the symmetry properties of the orbits under
the reflection at the $y$-axis $J (x, y) = (-x, y)$.
According to definitions (\ref{mdef1}) to (\ref{mdef3}) this reflection
acts on the Poincar\'e section as:
\begin{equation}
J(\chi, \psi, \epsilon) = (\chi, \psi, -\epsilon) \quad .
\end{equation}
Applying this reflection $J$ we can transform periodic orbits $\gamma_m$ of the
modular surface into periodic orbits $\gamma_b$ of the Artin billiard and vice
versa. Any prime periodic
orbit $\gamma_m$ of the modular surface is cut into segments
by its intersections with the $y$-axis and the boundary
of the fundamental domain. On these segments
it is a simple geodesic flow.
Applying $J$ to every second segment we get an orbit, that
fulfills the boundary conditions of the orbits for the Artin billiard.
Starting from a point $\underline{P}$ with positive
$x$-coordinate in the direction
$\underline{n}$ of $\gamma_m$ and transversing the
orbit $\gamma_m$, we thus construct an orbit $\tilde{\gamma}_b$
of the Artin billiard, which is periodic. The prime periodic part
of $\tilde{\gamma}_b$ is a unique function $G(\gamma_m)$ of the prime
periodic orbit $\gamma_m$. Only $\gamma_m$ and its mirror image
$J\hspace{1mm} \gamma_m$, but no
other prime periodic orbit of the modular surface,
yield the same orbit $G(\gamma_m) = G(J\hspace{1mm} \gamma_m)$.

Thus we have constructed a map $G$ from the set $M_m$ of prime
periodic orbits on the modular surface to the set $M_b$ of
prime periodic orbits in the Artin billiard. $M_m$ is divided into
symmetric $M^s_m = \{ \gamma_m \in M_m | J\hspace{1mm} \gamma_m = \gamma_m\}$
and asymmetric
$M^a_m = \{ \gamma_m \in M_m | J\hspace{1mm} \gamma_m \not= \gamma_m\}$
orbits.
The map $G$ is injective on $M^s_m$ and two-to-one on $M^a_m$.

In a similar geometric way we can construct the inverse of the map $G$.
Any prime periodic orbit $\gamma_b$ of the Artin billiard is cut into
segments by its reflections from the wall. Applying $J$ to every second
segment we get an orbit $\tilde{\gamma}_m$, which fulfills the boudary
conditions of the modular surface. Starting at $\underline{P}$ in the direction
$\underline{n}$ of $\gamma_b$ and
transversing $\gamma_b$ once, we thus construct an
orbit, which ends either at $(\underline{P}, \underline{n})$ or at
$(J\hspace{1mm}  \underline{P}, J\hspace{1mm} \underline{n})$,
depending on the number $n(\gamma_b)$ of intersections with the
unit circle:
\begin{enumerate}
\item $n(\gamma_b)$ even $\Rightarrow$
$\tilde{\gamma}_m$ ends at $(\underline{P}, \underline{n})$. Then
$\gamma_m := \tilde{\gamma}_m$
is a prime periodic orbit with the same length
$l$ and the same number $n$ of intersections with the
unit circle as $\gamma_b$. It is different
from its mirror image $J\hspace{1mm} \gamma_m$.
\item $n(\gamma_b)$ odd $\Rightarrow$
$\tilde{\gamma}_m$ ends at
$(J\hspace{1mm}  \underline{P}, J\hspace{1mm} \underline{n})$.
Then we create an orbit $\gamma_m$ by concatenating
$J\hspace{1mm} \tilde{\gamma}_m$ and
$\tilde{\gamma}_m$. This doubles the length:
$l(\gamma_m) = 2 \cdot l(\gamma_b)$
and the number $n$ of intersections with
the unit circle: $n(\gamma_m) = 2 \cdot n(\gamma_b)$.
The orbit $\gamma_m$ ends at
$(J^2 \hspace{1mm} \underline{P}, J^2\hspace{1mm}
\underline{n}) = (\underline{P}, \underline{n})$, thus it is
prime periodic. It is also symmetric: $\gamma_m = J\hspace{1mm} \gamma_m$.
\end{enumerate}
As there is no third possibility, all orbits $\gamma_b \in G(M^a_m)$ belong to
the first case, all orbits $\gamma_b \in G(M^s_m)$ to the second. It follows
that $G(M^a_m)$ and $G(M^s_m)$ are disjoint and that $G$ is surjective:
$M_b = G(M^a_m) \cup G(M^s_m)$.
\begin{enumerate}
\item $n(\gamma_b)$ even $\Leftrightarrow
\gamma_b \in G(M^a_m)$
\begin{eqnarray}
\Rightarrow G^{-1}(\{ \gamma_b \}) &=& \{\gamma_m, J\hspace{1mm} \gamma_m\} \\
l(\gamma_b) &=& l(\gamma_m)\\
n(\gamma_b) &=& n(\gamma_m)
\end{eqnarray}
\item $n(\gamma_b)$ odd $\Leftrightarrow
\gamma_b \in G(M^s_m)$
\begin{eqnarray}
\Rightarrow G^{-1}(\{ \gamma_b \}) &=& \{\gamma_m\} \\
l(\gamma_b) &=& l(\gamma_m)/2 \\
n(\gamma_b) &=& n(\gamma_m)/2
\end{eqnarray}
\end{enumerate}

\subsection{Periodic orbits of the Poincar\'e map and the Gau\ss{} map}
Let us denote the periodic orbits of the Poincar\'e map $\tilde{T}$ by
$\gamma_P$
and of the Gau\ss{} map $T_G$ by $\gamma_G$.
If the length of an periodic orbit $\gamma$ is $n$, then the
geometric length of the corresponding orbit of the geodesic flow is:
\begin{equation}
l(\gamma) = \sum_{i=1}^n r(\chi_i) \quad,
\end{equation}
where $r(\chi)$ was defined in
equation (\ref{mdefl}).
In analogy to the last section we divide the set $M_P$ of prime
periodic orbits of the Poincar\'e map $\tilde{T}$ into symmetric
$M^s_P = \{\gamma_P \in M_P | J\hspace{1mm} \gamma_P = \gamma_P\}$
and asymmetric
$M^a_p = \{\gamma_P \in M_P | J\hspace{1mm} \gamma_P \not= \gamma_P\}$ orbits.

As the first component $T_G(\chi)$ of the image
$\tilde{T}(\chi, \psi, \epsilon)$ does not depend on $\psi$ and $\epsilon$,
there is a trivial projection $(\chi, \psi, \epsilon) \mapsto \chi$,
which maps
any prime periodic orbit $\gamma_P$ of the Poincar\'e map $\tilde{T}$ onto
a periodic orbit $\gamma_G$ of the Gau\ss{} map $T_G$. The prime
periodic part of $\gamma_G$ is a function $H(\gamma_P)$ of the
prime periodic orbit $\gamma_P$.

On the other hand, if $\gamma_G$ is
a prime periodic orbit $\{\chi_i\}_{i \in {\bf Z}}$;
$\chi_{i + n} = \chi_i$; $\chi_{i+1} = T_G(\chi_i)$ of length $n(\gamma_G)$
of the
Gau\ss{} map, we can find corresponding orbits of the Poincar\'e map.
The coordinate $\psi_i$ is fixed uniquely by the periodic continued fraction
\begin{equation}
\psi_i = \frac{1}{[1/\chi_{i-1}] + \frac{1}{[1/\chi_{i-2}] + \cdots}}
\quad .
\end{equation}
Again we have to distinguish two cases:
\begin{enumerate}
\item $n(\gamma_G)$
even $\Rightarrow (\chi_0, \psi_0, 1)$ and $(\chi_0, \psi_0, -1)$
are starting points of two different prime periodic orbits
\begin{equation}
\{ \gamma_P, J\hspace{1mm} \gamma_P\} = H^{-1}(\{\gamma_G\})
\end{equation}
of length
\begin{equation}
n(\gamma_P) = n(J\hspace{1mm} \gamma_P) = n(\gamma_G)\quad .
\end{equation}
The corresponding geometric lengths are
also equal:
\begin{equation}
l(\gamma_P) = l(J\hspace{1mm} \gamma_P) = l(\gamma_G) \quad .
\end{equation}
\item $n(\gamma_G)$
odd $\Rightarrow (\chi_0, \psi_0, 1)$ and $(\chi_0, \psi_0, -1)$
belong to the same symmetric prime periodic orbit
\begin{equation}
\{ \gamma_P\} = H^{-1}(\{\gamma_G\})
\end{equation}
of length
\begin{equation}
n(\gamma_P) = 2 \cdot n(\gamma_G) \quad .
\end{equation}
The corresponding geometric
length, too, has doubled:
\begin{equation}
l(\gamma_P) = 2 \cdot l(\gamma_G) \quad .
\end{equation}
\end{enumerate}
Thus the set $M_G$ of prime periodic orbits of the Gau\ss{} map $T_G$
is split into two disjoint parts $H(M^a_P)$ and $H(M^s_P)$.

\subsection{Partition sums}
Now we have related the prime periodic orbits of the Artin billiard to
those of the modular surface, those to the ones
of the Poincar\'e map and finally the latter ones to
those of the Gau\ss{} map.
In order to determine dynamical zeta functions we have to
calculate sums over all prime periodic orbits of a quantity $\phi$, that
depends on the geometric length $l(\gamma)$ of the orbits. The sum
for the Artin billiard is:
\begin{eqnarray}
\sum_{\gamma_b \in M_b} \phi(l(\gamma_b))
&=& \sum_{\gamma_b \in G(M^a_m)} \phi(l(\gamma_b)) +
\sum_{\gamma_b \in G(M^s_m)} \phi(l(\gamma_b))
\nonumber \\
&=& \frac{1}{2}\cdot \sum_{\gamma_m \in M^a_m} \phi(l(\gamma_m)) +
\sum_{\gamma_m \in M^s_m} \phi(\frac{1}{2}\cdot l(\gamma_m))
\nonumber \\
&=& \frac{1}{2}\cdot \sum_{\gamma_P \in M^a_P} \phi(l(\gamma_P)) +
\sum_{\gamma_P \in M^s_P} \phi(\frac{1}{2}\cdot l(\gamma_P))
\nonumber \\
&=& \frac{1}{2}\cdot 2\cdot \sum_{\gamma_G \in H(M^a_P)} \phi(l(\gamma_G)) +
\sum_{\gamma_G \in H(M^s_P)} \phi(\frac{1}{2}\cdot 2 \cdot l(\gamma_G))
\nonumber \\
&=& \sum_{\gamma_G \in M_G)} \phi(l(\gamma_G))
\end{eqnarray}
equal to the sum for the Gau\ss{} map.

In the same way one shows, that
the number $n(\gamma_b)$ of intersections of an Artin billiard's orbit
with the unit circle is
equal to the length $n(\gamma_G)$ of the Gau\ss{} map's orbit:
\begin{enumerate}
\item $n(\gamma_G)$
even $\Rightarrow n(\gamma_G) = n(\gamma_P) = n(\gamma_m) = n(\gamma_b)$
\item $n(\gamma_G)$
odd $\Rightarrow n(\gamma_G) = n(\gamma_P)/2 = n(\gamma_m)/2 = n(\gamma_b)$
\end{enumerate}

Thus we can
calculate the dynamical zeta
functions for the Artin billiard as easily as those for the
Gau\ss{} map.

\subsection{Dynamical zeta functions}
When calculating the Ruelle zeta function $Z_R^{D}$
of a billiard with Dirichlet boundary condition we have to include the
phase factor $\exp(\pi i) = (-1)$ at every reflection from a wall
\cite{r22} and
thus the
phase factor $\exp(2\pi i) = 1$ at every reflection from a corner in
between two walls.
In the Artin billiard the $x$-component of the orbit's direction
changes its sign at every reflection from the lines $x=0$ or $x = 1/2$,
but it keeps it sign when reflected from the unit circle.
As the total number of sign changes along a periodic orbit must be even,
it is sufficent to include the phase factor $(-1)$
at every reflection from the unit circle, that is $n(\gamma_b) = n(\gamma_G)$
times.

The Ruelle zeta function $Z_R^{D}$ of the Artin billiard
with Dirichlet boundary condition is
defined for large enough real part $\Re \beta$
as a product over all prime periodic orbits $\gamma$:
\begin{equation}
\frac{1}{Z^{D}_R}(\beta) = \prod_{\gamma} (1 - \exp(- \beta \cdot l(\gamma))
\cdot (-1)^{n(\gamma)})
\end{equation}
This Euler product can be rewritten as a Dirichlet sum-like formula \cite{r6}:
\begin{eqnarray}
\frac{1}{Z^{D}_R}(\beta) &=& \exp(- \sum_{\gamma_b \in M_b} \sum_{m=1}^{\infty}
\frac{1}{m} \cdot (e^{-\beta l(\gamma_b)} \cdot (-1)^{n(\gamma_b)})^m) \\
&=& \exp(- \sum_{\gamma_G \in M_G} \sum_{m=1}^{\infty}
\frac{1}{m} \cdot (e^{-\beta l(\gamma_G)} \cdot (-1)^{n(\gamma_G)})^m) \\
&=& \exp(- \sum_{n=1}^{\infty} \frac{1}{n} \cdot
\sum_{x \in {\rm Fix} T_G^n} e^{-\beta l_n(x)} \cdot (-1)^n )
\label{mrue1} \quad ,
\end{eqnarray}
where ${\rm Fix} T_G^n$ is the set of fixpoints of $T_G^n$ and
\begin{equation}
l_n(x) := \sum_{i=0}^{n-1} r(T_G^i(x)) \quad .
\end{equation}
The sum of exponentials can be calculated by the transfer operator
method \cite{r6}. The transfer operator $L_{\beta}$
is the generalised Frobenius Perron operator
of the Gau\ss{} map:
\begin{eqnarray}
L_{\beta}f(x) &:=&
\sum_{\scriptsize \begin{array}{c}w:\\T_G(w) = x\end{array}}
\frac{f(w)}{|T_G^`(w)|^{\beta}} \\
&=&
\sum_{\scriptsize \begin{array}{c}w:\\T_G(w) = x\end{array}}
f(w) \cdot e^{-\beta r(x)} \quad.
\end{eqnarray}
The result is:
\begin{eqnarray}
\sum_{x \in {\rm Fix} T_G^n} e^{-\beta l_n(x)} &=&
\sum_{x \in {\rm Fix} T_G^n} \prod_{i = 0}^{n-1} \exp(-\beta r(T_G^i(x))
\\ &=& {\rm Tr}(L_{\beta}^n) - {\rm Tr}((-L_{\beta+1})^n) \quad .
\end{eqnarray}
Inserting this into equation (\ref{mrue1}) we get:
\begin{eqnarray}
\frac{1}{Z^{D}_R}(\beta) &=& \exp(- \sum_{n=1}^{\infty} \frac{1}{n} \cdot
(-1)^n \cdot
({\rm Tr}(L_{\beta}^n) - {\rm Tr}((-L_{\beta+1})^n))) \\
&=& \frac{\det(1 + L_{\beta})}{\det(1 - L_{\beta +1})}
\quad . \end{eqnarray}
The Selberg zeta function hence is given by:
\begin{eqnarray}
Z^{D}_S(\beta) &=& \prod_{k = 0}^{\infty} \frac{1}{Z^b_R(\beta +k)} \\
&=& \det(1 + L_{\beta}) \cdot \prod_{k = 1}^{\infty}
\frac{\det(1 + L_{\beta + k})}{\det(1 - L_{\beta + k})} \label{selD}
\quad . \end{eqnarray}
In the same way, only replacing the phase factor $(-1)$ by $1$, one
gets the Ruelle zeta function $Z^{N}_R$ of the Artin billiard with
Neumann boundary conditions:
\begin{equation}
\frac{1}{Z^{N}_R}(\beta) =
\frac{\det(1 - L_{\beta})}{\det(1 + L_{\beta +1})}
\quad . \end{equation}
The Selberg zeta function for Neumann boundary conditions then is:
\begin{equation}
Z^{N}_S(\beta) =
\det(1 - L_{\beta}) \cdot \prod_{k = 1}^{\infty}
\frac{\det(1 - L_{\beta + k})}{\det(1 + L_{\beta + k})} \label{selN}
\quad . \end{equation}
In \cite{r10} Bogomolny and Carioli conjectured, that the
Selberg zeta functions for the two cases of Dirichlet and
Neumann boundary conditions should be equal to $\det(1 + L_{\beta})$
and $\det(1 - L_{\beta})$. This conjecture is violated by the
additional factors in equations (\ref{selD}) and (\ref{selN}).

The poles and zeros on the line $\Re \beta = 1/2$,
which determine the spectrum of the Laplace-Beltrami operator \cite{r5},
are, however, not affected by these
additional factors: It follows from a result by
Pollicott in \cite{r21}, that
for $\Re \beta > 1$ all the eigenvalues
of $L_{\beta}$ have a modulus smaller than $1$. Thus $\det(1 + L_{\beta + k})$
and $\det(1 - L_{\beta + k})$ have no zeros for $\Re \beta = 1/2$
and $k \ge 1$.
In \cite{r17} it was shown, that $\beta = (1 - j) / 2$; $j \in {\bf N}_0$ are
the only poles of $L_{\beta}$. Thus $\det(1 + L_{\beta + k})$
and $\det(1 - L_{\beta + k})$ have no poles for $\Re \beta = 1/2$
and $k \ge 1$.

Finally, multiplying the Selberg zeta functions for Dirichlet and
Neumann boundary conditions, we get the Selberg zeta function $Z^{m}_S$ for
the modular surface \cite{r11} as:
\begin{eqnarray}
Z^{m}_S(\beta) &=& Z^{N}_S(\beta) \cdot Z^{D}_S(\beta) \\
&=& \det(1 - L_{\beta}) \cdot \det(1 + L_{\beta}) \quad .
\end{eqnarray}

\section{The Artin--Venkov--Zograf factorization formula}
In \cite{r15} Venkov and Zograf proved a remarkable factorization
formula for the Selberg zeta function for normal subgroups $\Gamma'$ of a
Fuchsian group $\Gamma$
with finite factor group $\Gamma / \Gamma'$. This
formula is well known in algebraic number theory and was found for
number theoretic zeta functions by Artin and Tagaki \cite{r18}. Starting
from the relation of the spectra of the free Schr\"odinger--operator on
the corresponding surfaces for the groups $\Gamma$ and $\Gamma'$
they showed that
the Selberg zeta function for the subgroup $\Gamma'$
can be simply expressed as a
product of zeta functions for $\Gamma$ with all finite dimensional
unitary representations of this group which have $\Gamma'$
in their kernel. To
be more precise, denote by $\chi^*(\Gamma/\Gamma')$ all inequivalent unitary
irreducible representations of the factor group
$\Gamma/\Gamma' = \{g \cdot \Gamma':
g \in \Gamma\}$, whose elements we denote by $\{ g \} $. If
$\tilde{\chi} \in \chi^*(\Gamma / \Gamma')$
then $\tilde{\chi}$ obviously defines
also a unitary representation $\chi$ of the group $\Gamma$ by
\begin{equation}
\chi(g) = \tilde{\chi}(g \cdot \Gamma')
\end{equation}
For $g \in \Gamma'$ this gives
\begin{equation}
\chi(g) = \tilde{\chi}(g \cdot \Gamma') = \tilde{\chi}(1 \cdot \Gamma') = {\bf
1}
\end{equation}
and hence $\Gamma' \in \mbox{ kernel } \chi$.

Consider next the generalized dynamical zeta
function $Z_S^{\Gamma}(\beta, \chi)$ of the
geodesic flow on the surface determined by the group $\Gamma$ \cite{r2}:
\begin{equation}
Z_S^{\Gamma}(\beta, \chi) := \prod_\gamma \prod_{k = 0}^\infty
\det\left(1 - \chi(P_\gamma) e^{- (\beta + k)l(\gamma)}\right) \quad,
\end{equation}
where $P_\gamma$ denotes an element of $\Gamma$ which
fixes the closed orbit $\gamma$, that means
$P_\gamma x_\infty = x_\infty$,
$P_\gamma x_{-\infty} = x_{-\infty}$, if $\gamma = (x_{-\infty}, x_{+\infty})$
is this orbit in the Poincar\'e half--plane ${\bf H}$
and $\chi$ is a representation of $\Gamma$ as discussed before.

Obviously for the trivial one dimensional representation $\tilde{\chi}_0$ with
$\tilde{\chi}_0(g \cdot \Gamma') \equiv 1$ we also have ${\chi}_0(g) = 1$
for all $g \in \Gamma$ and hence
\begin{equation}
Z_S^{\Gamma}(\beta, \chi_0) = Z_S^{\Gamma}(\beta)
= \prod_\gamma \prod_{k = 0}^\infty
\left(1- e^{- (\beta + k)l(\gamma)}\right)
\end{equation}
is the ordinary dynamical zeta function for the geodesic flow on the
surface ${\bf H}/\Gamma$ \cite{r1}, \cite{r2}.

The Artin--Venkov--Zograf formula then states \cite{r15}:
\begin{equation} \label{avz}
Z_S^{\Gamma'}(\beta) = \prod_{\tilde{\chi} \in \chi^*(\Gamma/\Gamma')}
Z_S^{\Gamma}(\beta, \chi)^{\dim \chi} \quad .
\end{equation}

The proof by the above authors uses the specific nature of the
algebraic structure of Fuchsian groups. However, it turns out
that this formula has an almost trivial interpretation when the
dynamics of the involved geodesic flows is taken into account.
This approach works at least for compact sufaces
of constant negative curvature and their finite sheeted
coverings. More interestingly, it works also for the modular surfaces,
that means the surface defined by the modular group $SL(2,{\bf Z})$ and its
finite coverings. It is expected that the same arguments
can be applied to general constant negative curvature surfaces
with finite volume, as soon as the thermodynamic formalism
approach to their dynamical zeta functions has been worked out
\cite{r11}. Interestingly enough, this approach even extends
to a case  strictly speaking not covered by the Venkov--Zograf
paper, namely the group $GL(2,{\bf Z})$ of all $2*2$ integer matrices with
determinant $\pm 1$ --- which is not a Fuchsian group.

We will show now, that the factorization of the dynamical
zeta function for the modular surface as dicussed in
the first sections of this paper can indeed be interpreted in
this way. To see this, remember the transfer operator
$\tilde{L}_\beta$ of the geodesic flow on the modular surface \cite{r11}:
\begin{equation}
\tilde{L}_\beta f(z,\epsilon) = \sum_{n = 1}^\infty
\left(\frac{1}{z+n}\right)^{2\beta}
f\left(\frac{1}{z+n}, -\epsilon\right) \quad .
\end{equation}
The group $SL(2,{\bf Z})$ is a normal subgroup of $GL(2,{\bf Z})$
and $GL(2,{\bf Z})/SL(2,{\bf Z})$
has just two elements which we
denote by $\{ g\}  = \pm 1$ corresponding to the two classes
$g \cdot SL(2,{\bf Z})$ with $g \in GL(2,{\bf Z})$ and
$\det g = \pm 1$. The group $GL(2,{\bf Z})/SL(2,{\bf Z})$ has just
two finite dimensional irreducible unitary
representations $\tilde{\chi}_1, \tilde{\chi}_2$, both one dimensional with
\begin{equation}
\tilde{\chi}_1(g \cdot SL(2,{\bf Z})) = 1
\end{equation} and \begin{equation}
\tilde{\chi}_2(g \cdot SL(2,{\bf Z})) = \det g
\end{equation}
for all $g \in GL(2,{\bf Z})$ \cite{r19}. The corresponding representations
$\chi_1, \chi_2$ of $GL(2,{\bf Z})$ with $SL(2,{\bf Z})$ in their kernel are
\begin{equation}
\chi_1(g) = 1
\end{equation} and \begin{equation}
\chi_2(g) = \det g \quad .
\end{equation}
The above transfer operator $\tilde{L}_\beta $ can
be rewritten also as follows:
\begin{eqnarray}
\tilde{L}_\beta f(z,\{ g\} )&=& \sum_{n = 1}^\infty
\left(\frac{1}{z+n}\right)^{2\beta}
f(\frac{1}{z+n}, \{ J Q T^{-n}\} \{ g\} )\label{lb1}\\
&=& \sum_{n = 1}^\infty
\left(\frac{1}{z+n}\right)^{2\beta}
\tilde{\chi}_L(\{ T^n Q J\} ) f(\frac{1}{z+n}, \{ g\} )
\label{lb}
\end{eqnarray}
where
\begin{equation}
\tilde{\chi}_L : GL(2,{\bf Z})/SL(2,{\bf Z}) \to
\mbox{Aut}({\bf C}(GL(2,{\bf Z})/SL(2,{\bf Z})))
\end{equation}
denotes the so called left--regular representation
of the group $GL(2,{\bf Z})/SL(2,{\bf Z})$
on the two dimensional space of complex valued functions on the
group $GL(2,{\bf Z})/SL(2,{\bf Z})$ defined quite generally for any finite
group
$G$ as \cite{r19}:
\begin{equation}
\tilde{\chi}_L(g') f(g) = f({g'}^{-1} g) \mbox{ for } g,g' \in G
\end{equation}
and $f \in {\bf C}(G)$, the space of complex functions on $G$
with $\dim {\bf C}(G) = \mbox{ order of } G$.

In equation (\ref{lb1}) $J$ denotes the reflection $Jz = -z^{\ast}$
and $T$ and $Q$ are the generators $Tz = z+1$ resp. $Qz = -1/z$ of
the group $PSL(2,{\bf Z})$. They are given by the corresponding
matrices in $GL(2, {\bf Z})$.
We also made use of the fact that
$\{ J Q T^{-n}\} \{ g\}  = -\{ g\} $ for all $n$ and all $\{ g\}  = \pm 1$
since
$\det(JQT^{-n}) = -1$ for all $n$.

To understand expression (\ref{lb}) better remember that
the Poincar\'e map for the Artin-billiard was simply
\begin{equation}
P(x,y) = \left(T_G x, \frac{1}{y + [1/x]}\right)\quad .
\end{equation}
Consider then a geodesic on the Artin billiard starting in the
point $x, y$ of the
Poincar\'e section. On the modular surface this point is
covered by
the two points $(x, y, \{ g\} )$ with
$\{ g\}  = \pm 1 \in GL(2,{\bf Z})/SL(2,{\bf Z})$.
When starting in one of them, say
$(x, y, 1)$, the point will come back to a lift of the
points $P(x,y)$ in the Poincar\'e section of the Artin billiard.
{}From the symbolic dynamics of Series et al. \cite{r12}, \cite{r13} it
follows that the Poincar\'e map $P$ just corresponds to the map
$JGT^{-n}$ of the endpoints $x_{-\infty}$ and $x_{+\infty}$
of the corresponding half circle $\gamma$ in the upper half plane
and hence the geodesic arrives in the point
$(P(x,y), \{ JQT^{-n}\}) = (P(x,y),-1)$,
which then defines the Poincar\'e
map for the geodesic flow on the modular surface.
The left regular representation $\tilde{\chi}_L$ can be decomposed completely
into its irreducible components \cite{r19} which are just
all the finite--dimensional irreducible unitary representations
$\chi^*(GL(2,{\bf Z})/SL(2,{\bf Z}))$
and therfore $\tilde{\chi}_1$ and $\tilde{\chi}_2$. It is known
\cite{r19} that each of these representations occurs in
$\tilde{\chi}_L$ just as many times as given by its dimension, hence
exactly once. The transfer operator $\tilde{L}_\beta$ hence can
be written as
\begin{eqnarray}
\tilde{L}_\beta &=& L_{\beta, \chi_1} \oplus L_{\beta, \chi_2} \\
\mbox{with } L_{\beta, \chi_1} f(z) &=& \sum_{n = 1}^\infty
\left(\frac{1}{z+n}\right)^{2\beta} f(\frac{1}{z+n}) \\
\mbox{and } L_{\beta, \chi_2} f(z) &=& -\sum_{n = 1}^\infty
\left(\frac{1}{z+n}\right)^{2\beta} f(\frac{1}{z+n})
= -L_{\beta, \chi_1} f(z)\\
\mbox{since } \chi_2(T^n QJ) &=& -1 \mbox{ for all }n \quad .
\end{eqnarray}

It is then straightforward to show, that the zeta functions for the
Artin billiard with Neumann and Dirichlet boundary conditions
coincide with the zeta functions for $GL(2,{\bf Z})$ with representations
$\chi_1$ and $\chi_2$:
\begin{eqnarray}
Z_S^N(\beta) &=& Z_S^{GL(2,{\bf Z})}(\beta, \chi_1)\\
\mbox{and } Z_S^D(\beta) &=& Z_S^{GL(2,{\bf Z})}(\beta, \chi_2)
\end{eqnarray}
and our factorization
\begin{equation}
Z_S^{SL(2,{\bf Z})}(\beta) = Z_S^D(\beta) \cdot Z_S^N(\beta)
\end{equation}
coincides indeed with the Venkov--Zograf formula
for $SL(2,{\bf Z})$ when considered a subgroup of $GL(2,{\bf Z})$.

The above sequence of arguments extends immediately to any finite
sheeted covering of the modular surface. If $\Gamma'$ denotes
the corresponding subgroup of the group $\Gamma = SL(2,{\bf Z})$, then the
covering group of ${\bf H}/\Gamma'$ with
respect to ${\bf H}/\Gamma$
is just $\Gamma/\Gamma'$. The Poincar\'e map for the geodesic flow on
the covering surface can be constructed from the Poincar\'e
map $\tilde T:I_2 \times {\bf Z}_2 \to I_2 \times {\bf Z}_2$ (\ref{ttilde})
of the modular surface in complete analogy to our
procedure in going from the Artin billiard to the modular
surface: in the present case every point in the Poincar\'e
section of the geodesic flow on the modular surface is
covered by $d = [\Gamma:\Gamma'] = \#\{\{ g\}  \in \Gamma/\Gamma'\}$
points given by $(x,y,\epsilon,\{ g\} ), \{ g\}  \in \Gamma/\Gamma'$.
A geodesic starting at the point $(x, y, \epsilon)$ in the
Poincar\'e section of the modular surface and
arriving at $\tilde T(x,y,\epsilon)$ gets therefore lifted
to a geodesic on the covering surface starting in
$(x,y,\epsilon,\{ g\} )$ and arriving at the point
$(\tilde T(x,y,\epsilon), \{ QT^{-n\epsilon}\} \{ g\} )$ in the lifted
Poincar\'e section, where $n = [1/x]$.
This gives the following Poincar\'e map:
\begin{equation}
\tilde T^{\Gamma'}(x, y, \epsilon, \{ g\} )
= (\tilde T(x, y, \epsilon), \{ Q T^{-n\epsilon}\} \{ g\} ), \quad n = [1/x]
\quad .
\end{equation}
The generalized transfer operator hence has the form:
\begin{equation}
\tilde{L}_\beta^{\Gamma'} f(z, \epsilon, \{ g\} ) =
\sum_{n=1}^{\infty} \left(\frac{1}{z+n}\right)^{2\beta}
f\left(\frac{1}{z+n}, -\epsilon, \{ T^{n\epsilon}Q\} \{ g\} \right) \quad .
\end{equation}
Introducing again the left regular representation
$\tilde{\chi}_L$ of the group $\Gamma/\Gamma'$, on the space of
complex functions on
$\Gamma/\Gamma'$ we can write $\tilde{L}_\beta^{\Gamma'}$ also as
\begin{equation}
\tilde{L}_\beta^{\Gamma'} f(z, \epsilon, \{ g\} ) =
\sum_{n=1}^{\infty} \left(\frac{1}{z+n}\right)^{2\beta}
\tilde{\chi}_L(\{ QT^{-n\epsilon}\} )
f\left(\frac{1}{z+n}, -\epsilon, \{ g\} \right) \quad .
\end{equation}
The left regular representation $\tilde{\chi}_L$ decomposes again as
\cite{r19}:
\begin{equation}
\tilde{\chi}_L = \bigoplus_{\chi_i \in \chi^*(\Gamma/\Gamma')}
n_i \tilde{\chi}_i
\end{equation}
with $n_i$ being the dimension of the irreducible representation
$\tilde{\chi}_i$ of $\Gamma/\Gamma'$.
This shows that the transfer operator $\tilde{L}_\beta^{\Gamma'}$ for
the geodesic flow on the covering surface ${\bf H}/\Gamma'$
of the modular surface can be rewritten in the appropriate basis
of the space of complex functions on $\Gamma/\Gamma'$ as
\begin{equation}
\tilde{L}_\beta^{\Gamma'} \underline{f}(z, \epsilon)  =
\sum_{n=1}^{\infty} \left(\frac{1}{z+n}\right)^{2\beta} \cdot
\end{equation}\begin{displaymath}
\cdot \left( \begin{array}{rll}
\begin{array}{rl}
\tilde{\chi}_1(QT^{-n\epsilon}) &\\
\ddots &\\
&\tilde{\chi}_1(QT^{-n\epsilon}) \\
\end{array}
& \Bigg\}  n_1\mbox{--times} & 0
\\
\ddots &&
\\
0 \qquad \qquad \qquad &
\begin{array}{rl}
\tilde{\chi}_k(QT^{-n\epsilon}) &\\
\ddots &\\
&\tilde{\chi}_k(QT^{-n\epsilon}) \\
\end{array}
& \Bigg\}  n_k\mbox{--times}
\end{array} \right)
\underline{f}\left(\frac{1}{z+n}, -\epsilon\right)
\end{displaymath}
But this shows that
\begin{equation}
\det(1 - \tilde{L}_\beta^{\Gamma'}) = \prod_{i=1}^k
\det\left(1 - \tilde{L}_{\beta,\chi_i}^{\Gamma} \right)^{n_i}
\end{equation}
where
\begin{equation}
\tilde{L}_{\beta,\chi_i}^{\Gamma} \underline{f}(z, \epsilon)
= \sum_{n=1}^{\infty} \left(\frac{1}{z+n}\right)^{2\beta}
\chi_i(QT^{-n\epsilon}) \underline{f}\left(\frac{1}{z+n}, -\epsilon\right)
\end{equation}
is the transfer operator for the geodesic flow on
the modular surface with unitary representation $\chi_i$
of the group $\Gamma = SL(2,{\bf Z})$ with $\Gamma'$ in its kernel.
{}From this the Venkov--Zograf formula (\ref{avz}) follows immediately, since
\begin{equation}
Z^{\Gamma}_S(\beta, \chi_i) = \det(1 - \tilde{L}^{\Gamma}_{\beta, \chi_i})
\end{equation}
as can be verified by standard arguments.

Of special interest \cite{r16} is the case of the principal congruence
subgroup $\Gamma(2)$ of the modular group with
\begin{equation}
\Gamma(2) := \{g \in SL(2,{\bf Z}): g \mbox{ mod } 2 = {\bf 1} \} \quad .
\end{equation}
Since
\begin{equation}
SL(2,{\bf Z}) / \Gamma(2) \cong SL(2, {\bf Z}_2) \cong S_3 \quad,
\end{equation}
where $S_3$ is the group of permutations of three elements,
${\bf H}/ \Gamma(2)$ is a sixfold covering of the modular
surface. Indeed, its fundamental domain can be chosen as \cite{r3}:
\begin{equation}
\overline{F}' = \{ z \in {\bf H}: 0 \le |\Re z| \le 1,
|z \pm 1/2| \ge 1/2\} \quad .
\end{equation}
Its zeta function $Z_S^{\Gamma(2)}(\beta)$ can be written as
\begin{equation}
Z_S^{\Gamma(2)}(\beta) =
Z_S^{SL(2,{\bf Z})}(\beta, \chi_1) \cdot Z_S^{SL(2,{\bf Z})}(\beta, \chi_2)
\cdot \left(Z_S^{SL(2,{\bf Z})}(\beta, \chi_3)\right)^2
\end{equation}
where $\tilde{\chi}_1$ denotes the trivial representation of $S_3$,
$\tilde{\chi}_2$ is the one dimensional representation with
$\tilde{\chi}_2(\tau) = \pm 1$ if $\tau$ is an even resp. odd permutation.
The two dimensional representation $\tilde{\chi}_3$
is given by the rotation of the plane
by the angle $\pm 2\pi /3$ respectively the reflection on the
$y$--axis: $x \to -x$ and $y \to y$ \cite{r19}.
Since $\pm Q$ and $\pm T$ generate $SL(2,{\bf Z})$ any of its
representations is uniquely determined by giving the
representation of $Q$ and $T$. For the case $\Gamma(2)$ we need the
representations which have $\Gamma(2)$ in their
kernels and these are the following:
\begin{eqnarray}
\chi_1(g) &=& 1 \mbox{ for all } g \in SL(2,{\bf Z}) \\
\chi_2(Q) &=& -1 \\
\chi_2(T) &=& -1 \\
\chi_3(Q) &=&
\left(\begin{array}{rr}
-\cos 2\pi /3 & \sin 2\pi /3 \\
\sin 2\pi /3 & \cos 2\pi /3 \\
\end{array}\right) \\
\chi_3(T) &=&
\left(\begin{array}{rr}
-1 & 0 \\ 0 & 1
\end{array}\right) \quad ,
\end{eqnarray}
where we have, according to (\ref{psl}), identified
$Q = \left(\begin{array}{rr} 0 & -1 \\ 1 & 0 \end{array}\right)$
and
$T = \left(\begin{array}{rr} 1 & 1 \\ 0 & 1 \end{array}\right)$
with matrices in $SL(2,{\bf Z})$.

The Venkov--Zograf formula then allows to determine the spectrum of the free
Schr\"odinger operator on the surface ${\bf H}/\Gamma(2)$ through
the spectra of the Schr\"odinger operators on the modular
surface with the boundary conditions
\begin{equation}
\underline{f}(gz) = \chi_i(g) \underline{f}(z)
\end{equation}
with $g$ the two generators $T$ and $Q$, what one expects
also from purely spectral reasons. These spectra are again
determined by the nontrivial zeros of $Z_S^{\Gamma'}(\beta, \chi_i)$
on the line $\Re \beta = 1/2$ and hence by those $\beta$-values on this
line for which $\tilde{L}_{\beta, \chi_i}$ has $\lambda = 1$ in its
spectrum.

\pagebreak
\begin{figure} \label{fullview}
\vspace{.1cm}
\caption{The shaded area is the domain of the Artin billiard. It is
half as big as the fundamental domain of the modular surface.
The geodesic flow takes place along vertical lines or halfcircles
centered on the $x$-axis. Examples are the halfcircles $\gamma_1$,
$\gamma_2$ and its reflection image $J \gamma_2$.}
Comment for users of the automated preprint bulletin board:
This figure has not been added to the file, as it is purely
introductory. Pictures of the symbolic plane can be found for
example in the book by Gutzwiller or Terras.
\end{figure}
{}.

\newpage
\tableofcontents

\end{document}